\documentstyle[aaspp4]{article}
\lefthead{FRYER, COLGATE, \& PINTO}
\righthead{Pulsar in 1987A}

\begin{document}

\title{Iron Opacity and the Pulsar of Supernova 1987A}
\author{Chris L. Fryer}
\affil{Lick Observatory, University of California Observatories,\\
       Santa Cruz, CA\ \ \ 95064\\
       cfryer@ucolick.org}
\authoremail{cfryer@ucolick.org}

\author{Stirling A. Colgate}
\affil{Los Alamos National Laboratory,\\
       MS B275, Los Alamos, NM\ \ \ 87545\\
       colgate@lanl.gov}
\authoremail{colgate@lanl.gov}
\and
\author{Philip A. Pinto}
\affil{Steward Observatory, University of Arizona,\\
       Tucson, AZ\ \ \ 85721\\
       pinto@as.arizona.edu}
\authoremail{pinto@as.arizona.edu}

\begin{abstract}

Neutron stars formed in Type II supernovae are likely to be initially
obscured by late-time fallback.  Although much of
the late-time fallback is quickly accreted via neutrino cooling, some
material remains on the neutron star, forming an atmosphere which slowly
accretes through photon emission.  In this paper, we derive structure
equations of the fallback atmosphere and present results of
one-dimensional simulations of that fallback. The atmosphere remaining
after neutrino cooling ($L_{\nu}$) becomes unimportant 
($L_{\nu} \lesssim L_{\rm Edd,e^-}$, the Compton Eddington limit) is only a
fraction of the total mass accreted ($\lesssim 10^{-8} M_{\rm acc} =
10^{-9}M_\odot$).  Recombined iron dominates the opacity in the outer regions
leading to an opacity $10^3-10^4$ times higher than that of electron scattering
alone. The resultant photon emission of the remnant atmosphere is
limited to $\lesssim 10^{-3} L_{\rm Edd,e^-}$.  
The late-time evolution of this
system leads to the formation of a photon-driven wind from the accretion of
the inner portion of the atmosphere, leaving, for most cases, a bare
neutron star on timescales shorter than a year.  The degenerate remnant of
1987a may not be a black hole.  Instead, the fallback material may have
already accreted or blown off in the accretion-driven wind.  If the neutron
star has either a low magnetic field or a low rotational spin frequency, we
would not expect to see the neutron star remnant of 1987a.
\end{abstract}


\section{Introduction}

The mechanism for Type II supernovae (SNe) is triggered by the 
collapse of massive stars (Burbidge et al. 1957).  As the stellar core
collapses its gravitational energy is released through the emission of
neutrinos (Colgate \& White 1966; Bethe 1990). 
A small fraction of these neutrinos are absorbed and the heating
this provides drives the supernova explosion.  Prior to the appearance of
supernova 1987A, this mechanism, though well-developed, had no direct
observational validation.  The observation of neutrinos from SN 1987a at
both the IMB (Bionta et al. 1987) and Kamiokande (Hirata et al. 1987) 
detectors provided the first such validation.  
Both the $\sim 2$s duration of the
neutrino burst and the number flux of neutrinos agree well with the
predictions of the core-collapse model.  However, with all the reassurances
SN 1987A provides in support of the standard model, it also brings many
new puzzles.  One such unsolved mystery is the lack of detection of the
neutron star formed during the collapse.

Neutron stars formed in type II supernovae are thought to emit radiation as
pulsars, the Crab pulsar (Rickett \& Seiradakis 1982) being the canonical
example.  Even if the pulsar were not directed along our line of sight, its
radiation would heat the surrounding supernova remnant and add to its
bolometric luminosity.  At present, all of the luminosity of SN 1987A can
be accounted for by a radioactive decay model (with corrections for the
``freeze-out'' of the ionization state at late times) consistent with the
production of $0.075M_\odot$ of $^{56}$Ni (Kozma \& Fransson 1998).  The
current lack of evidence of any ``additional'' energy source places a limit
on any emission from the pulsar in SN 1987A at $\sim 10^{37}$erg s$^{-1}$
(Suntzeff et al. 1992, Kozma \& Fransson 1998), a factor of 50 times lower
than that of the Crab pulsar.

Where, then, is the neutron star formed by SN 1987a?  First, it may not be a
pulsar.  SN 1987A's remnant may have a magnetic field which is too weak, or
it may not be spinning sufficiently rapidly, to drive a strong pulsar like
that in the Crab.  Alternatively, the neutron star could obscured by 
material which is falling back onto the neutron star.   Even so,
one would expect accretion to provide a sufficient energy source to have
been observed (Woosley, Hartman, \& Pinto 1989). Houck \& Chevalier (1991)
have shown that the ever-present fallback of matter onto the neutron star
(Woosley \& Weaver 1995) would produce a luminosity roughly equal to the
Compton Eddington Luminosity ($L_{\rm Edd,e^-}=4 \times 10^{38}$erg
s$^{-1}$), over an order of magnitude greater than the observed limit 
of $10^{37}$erg s$^{-1}$.

It would seem we are forced to accept that the neutron star in SN 1987A
must have collapsed further into a black hole whose accretion luminosity
might be below the observational limits.  Although the duration of the
neutrino burst from 1987A precludes further collapse during the first $\sim
20$ seconds, the neutron star may have collapsed into a black hole at some
later time due either to the accretion of a sufficient fall-back mass and/or
to some change in its equation of state.

The late-time fallback calculated by Woosley \& Weaver (1995) could,
presumably, push the neutron star beyond the critical mass, forcing it to
collapse into a black hole.  However, their calculations estimate typical
fallback masses given the likely SN 1987A progenitors to be $\sim
0.2M_\odot$ and total neutron star remnant gravitational 
masses $\lesssim 1.7 M_\odot$.  In fact, because 
$0.07M_\odot$ of $^{56}$Ni was certainly ejected
from SN 1987a, the maximum mass of its compact remnant cannot be greater
than this value or it would have accreted all of the $^{56}$Ni production.
If this low-mass neutron star did indeed collapse into a black hole, many
of the current equations of state for dense matter must be incorrect.
Bethe \& Brown (1995) have used the collapse of the neutron star initially
formed in SN 1987A to argue for an equation of state softened by pion
condensates.  Such an exotic equation of state, or something akin to it, is
required to explain such a low stable mass for neutron stars.

The news of the demise of 1987A's neutron star may be premature, however.
In this paper, we study the characteristics of the fallback material to
question the high accretion luminosity derived by Houck \& Chevalier
(1992).  Their result is based on the assumption that the opacity of the
fallback material is comparable to Compton scattering.  We show in this paper
that for conditions appropriate to many fallback scenarios, the actual
opacity can be more than 3-4 orders of magnitude greater than that due to
Compton scattering alone, both in the first few days past explosion and for
many years thereafter.  The appropriate ``Eddington limit'' for the
accreting neutron star is thus over three orders of magnitude lower than that
employed by Houck \& Chevalier (1992) and 
would likely remain undetectable in the
lightcurve of SN 1987A for centuries to come.  Indeed, in our simulations
the accreting material drives a photon wind which halts further accretion,
leaving a bare neutron star on timescales of less than a year.  If the
compact remnant of 1987A has not yet formed a strong pulsar, it may
still be a neutron star.

We first calculate the effect of this high opacity on the evolution of an
atmosphere formed by the fallback process.  In \S 2, we discuss the 
nature and effects of the fallback which occurs during the supernova 
explosion as the shock progresses through the hydrogen envelope, 
sending a reverse shock back toward the nascent neutron star and 
driving very rapid accretion $\dot{M}>10^4 M_\odot \, {\rm yr^{-1}}$ 
(Woosley \& Weaver 1995).  The structure of this atmosphere
is derived in \S 2.1 and compared (favorably) to the numerical calculations
in \S2.2.  Most of the atmosphere is accreted via neutrino emission 
leaving an atmosphere which cools through photon diffusion.  However, 
the opacity of the outer layers of this atmosphere is 3-4 orders 
of magnitude higher than that of electron scattering and is 
dominated by recombined heavy elements.  These outer ``blanket'' 
layers are blown off as a wind driven by the flux diffusing from the 
inner, hotter (dissociated), and consequently lower-opacity
layers at the Compton Eddington rate(\S 2.3).  
This energy flux, derived from the gravitational energy of the
accreting matter, goes primarily into reducing the gravitational binding
energy of the ejected matter and produces a photon flux nearly 3
orders of magnitude below the detection limit.  This process continues 
until all the matter in the atmosphere is either accreted or ejected.

Although numerical considerations limit our calculations to only a 
small fraction of the duration of this process, the 
outcome of the process is not in doubt. In a relatively short time, hours
to weeks after neutrino cooling has shut down, the photon heat flow at
$L_{\rm Edd,e^-}$ at the base of the atmosphere is sufficient to
remove all the mass of the initial atmosphere.  
Further accretion could occur at later times (say 10 years after 
the supernova event) and understanding its fate and the luminosity 
it produces requires a more detailed discussion of line driven 
opacity, which we relegate to \S 3.  Here again, we find the 
accretion luminosity to be much less than the Compton Eddington 
rate.  Line-driven opacity dominates the opacity in all fallback 
accretion scenarios, and the resultant luminosities are below 
the current detection threshold.  A neutron star may yet 
lurk in the ashes of SN 1987A.

\section{Supernova Fallback}

The first calculations of fallback in supernovae occurred soon after the
initial proposal of the neutrino-driven supernova mechanism (Colgate 1971,
Bisnovatyi-Kogan \& Lamzin 1984).  In these calculations, the fallback 
results from the loss in pressure support of the expanding ejecta as the
material near the neutron star quickly cooled through neutrino emission.
Lacking pressure support, the matter falls back onto the neutron star
because of the strong gravitational field.  In this mechanism, material
began to fall back onto the neutron star almost immediately after the
launch of the supernova explosion.  Since this time, a new mechanism
causing fallback at later times ($10-10^4$s after the explosion) was
discovered through simulations of 1987A supernova explosions (Shigeyama 
et al. 1988; Woosley 1989).  This fallback arises from a
deceleration of the inner layers of the explosion. As the initial
explosion shock encounters the massive envelope of the star, material piles
up behind it, sending a pressure wave (a shock in most simulations) back
toward the center and decelerating the inner layers, driving material back
onto the nascent neutron star (Herant \& Woosley 1994; Woosley \& Weaver
1995).

Figure 1 shows the rate of mass infall from fallback for three different
supernova models calculated from the simulations of Woosley \& Weaver
(1995).  Note that the bulk of the accretion occurs at very rapid infall
rates ($\gtrsim 10^4M_\odot\,$y$^{-1}$).  The fate of fallback material with these
infall rates has been studied in detail (Chevalier 1993, 1996; Houck \&
Chevalier 1992; Brown 1995; Fryer, Benz, \& Herant 1996).  Above a critical
infall rate $\sim 10^{-4}M_\odot\,$y$^{-1}$, this material cools quickly through
neutrino emission and accretes onto the neutron star.  Below this, photon
radiation dominates the cooling of the infalling material and the photon
Eddington limit becomes important.  All of these calculations, however,
assume a constant infall rate which is clearly not appropriate at later
times.

In this section we are interested in the structure of the atmosphere which
remains around the neutron star after neutrino cooling becomes inefficient
and, hence, must accrete through photon cooling.  This material is
comprised of the last bit of fallback onto the neutron star at times when
the infall rate is decreasing rapidly (see Figure 1).  We
derive the structure equations of the atmosphere for a variable infall
rate, ultimately deriving not only the density and temperature profiles but
also the amount of mass that piles on top of the neutron star.  We compare
these semi-analytic results to those of detailed hydrodynamical
simulations.

\subsection{Structure Equations}

The characteristics of an accretion shock for a constant  infall
rate are well known (e.g. Chevalier 1993).  In the  strong shock
limit for a radiation-pressure dominated  ($\gamma = 4/3$) gas,
the pressure ($P_{\rm sh}$), density  ($\rho_{\rm sh}$), and
hence, entropy ($S_{\rm sh}$) at the  shock are given by:
\begin{equation}
P_{\rm sh}=7/6 \rho_{\rm ff} V^2_{\rm ff},
\end{equation}
\begin{equation}
\rho_{\rm sh}=7 \rho_{\rm ff},
\end{equation}
and
\begin{equation}
S_{\rm sh}=P^{3/4}_{\rm sh}/\rho_{\rm sh},
\end{equation}
where $V_{\rm ff}= \sqrt{2 G M_{\rm NS}/R_{\rm
sh}}$ is the free-fall velocity, $\rho_{\rm ff}=\frac{\dot{M}} {4 \pi
R_{\rm sh}^2 V_{\rm ff}}$ is the density of the accreting matter, $G$ is
the gravitational constant, $M_{\rm NS}$ is the neutron star mass, and
$R_{\rm sh}$ is the shock radius.  We parameterize the variable accetion
rate as an exponential decrease in time, $\dot{M} \equiv \dot{M}_0
(t/t_0)^{-\alpha}$.  By setting $t=t_{\rm ff}$, we can write the infall
rate as a function of radius:
\begin{equation}
\dot{M}=\dot{M}_0 \left( \frac{2}{\pi} G^{1/2} M^{1/2}_{\rm NS}
t_0 \right)^{\alpha} R_{\rm sh}^{-3/2 \alpha}.
\end{equation} 
Using these equations, we can express the entropy
of the shock as a function of radius:
\begin{equation}\label{eq:entr}
S_{\rm sh}=3.2 \times10^{-18} \dot{M}_0^{-1/4} M_{\rm NS}^{(7-\alpha)/8}
\left( \frac{6.1\times10^3}{t_0} \right)^{\alpha/4}  
R_{\rm sh}^{3/8(\alpha-1)}=S_0  R_{\rm sh,6}^{3/8(\alpha-1)}
\end{equation} 
where $\dot{M}_0,M_{\rm NS}$, and $t_0$ are all given in cgs units and
$R_{\rm sh,6}$ is the shock radius in units of $10^6$cm.  All previous work
assumed a constant accretion rate ($\alpha=0$) and the resulting atmosphere
was convectively unstable ($\partial S / \partial R = -\frac{3}{8} S_0
R_{sh}^{-11/8} < 0$).

Convection will strive to flatten a negative entropy  profile,
and for the cases where $\alpha < 1$,  the entropy of the
resulting accretion atmosphere (within the shock) is constant.
By assuming  pressure equilibrium for these constant entropy
atmospheres and a neutron star mass much greater  than the mass
of the surrounding atmosphere,  the pressure gradient is given by
(see, for  example, Colgate, Herant, \& Benz 1993):
\begin{equation}\label{eq:equi} 
dP/dR=-g\rho=-M_{\rm NS} G \rho /R^2.
\end{equation} 
Since the entropy is independent of radius, we can
integrate equation (\ref{eq:equi}) to obtain:
\begin{equation} 
P=\left[ \frac{1}{4} M_{\rm NS} G S_{0}^{-1} 
(1/R-1/R_{\rm out}) +  P_{\rm out}^{1/4} \right]^4,
\end{equation} 
where $R_{\rm out}$ and $P_{\rm out}$ are the radius and pressure of the
outer boundary of the atmosphere.  Since we define the atmosphere as the
material within the accretion shock, $R_{\rm out}$ is equivalent to the
accretion shock radius and $P_{\rm out}$ is the pressure at that radius.
Using this relation for a radiation-dominated gas of pressure $P=1/3 a T^4$
and entropy $S=4/3 a T^3/ \rho$, one can derive the structure equations for
the atmosphere (Colgate, Herant, \& Benz 1993)\footnote{Our derivation
differs from the work of Colgate, Herant, \& Benz (1993) in that we assume
we can neglect the effects of electron-positron pairs.  For the
temperatures involved in this work, this assumption is valid.} including
the density and temperature profiles as well as the atmosphere mass.

For the fallback scenarios we consider, as the last bit of material falls
onto the neutron star, the fallback rate decreases dramatically: $\alpha >
1$ (see Figure 1).  Although convection will not flatten the entropy
profile for these systems, we can still derive the structure equations for
these atmospheres assuming that as the shock moves outward, the entropy
profile is maintained.  Then, again by assuming pressure equilibrium, we
can derive the pressure profile of the atmosphere:
\begin{equation} 
P=\left[ \frac{1}{4}M_{\rm NS}G S_{0}^{-1}
\left(R^{-5/8-3\alpha/8}-R_{\rm out}^{-5/8-3\alpha/8}\right) 
+ P_{\rm out}^{1/4}\right]^4.
\end{equation}
$R_{\rm out}$ is generally much beyond the regions that we
consider ($R_{\rm out} \approx R_{\rm Bondi-Hoyle}$) and hence it
and the outer pressure can be neglected.  The structure
equations then become:
\begin{equation}\label{eq:pres} 
P=\left[\frac{1}{4}M_{\rm NS}G S_{0}^{-1} \right]^4 R^{-5/2-3\alpha/2}= P_0
R_6^{-5/2-3\alpha/2} \; {\rm g \, cm^{-1} \, s^{-2}},
\end{equation}
\begin{equation}\label{eq:temp}
T=\left(\frac{3}{a}\right)^{1/4}\left[\frac{1}{4}M_{\rm NS}
G S_{0}^{-1} \right]  R^{-5/8-3\alpha/8}=T_0
R_6^{-5/8-3\alpha/8}\; {\rm K},
\end{equation}
\begin{equation}\label{eq:dens}
\rho=S_0^{-1}\left[\frac{1}{4}M_{\rm NS}
G S_{0}^{-1} \right]^3  R^{-3/2(1+\alpha)}=\rho_0
R_6^{-3/2(1+\alpha)}\; {\rm g \, cm^{-3}},
\end{equation}
and
\begin{equation}\label{eq:mass} 
M_{\rm encl}=\frac{4\pi}{3/2(1-\alpha)} 
S_0^{-1}\left[\frac{1} {4}M_{\rm NS}G S_{0}^{-1}\right]^3 
\left[R^{3/2(1-\alpha)}_{\rm NS}-
R^{3/2(1-\alpha)}_{\rm max}\right] M_{\odot},
\end{equation} 
where $R_{\rm NS}$ is the radius of the neutron star and $R_{6}$ is the
radius in units of $10^6$cm. In next section, we compare this
analytic derivation of the atmosphere's structure to those calculated by
numerical simulation.  These analytic results will give us a basis from
which we can build our understanding of the evolution of these atmospheres.

\subsection{Comparison to Simulations and the Atmosphere Mass}

The structure equations do not uniquely define the atmosphere around the
neutron star. They depend both upon the parameter $\alpha$ and the initial
accretion rate $\dot{M}_0 t_0^{\alpha}$.  Although the uncertainty in
fallback rates allows relative freedom in the choice of $\alpha$, we can
constrain $\dot{M}_0 t_0^{\alpha}$ by insisting that we consider only the
mass of the atmosphere which is not cooled primarily by neutrinos.  The
neutrino cooling of the atmosphere at temperatures $\lesssim 1$ MeV is
dominated by electron capture on protons and can be approximated by (Bethe
1990):
\begin{equation} 
L_{\nu}=\int_{r=R_{\rm NS}}^{\infty} 2 \times 10^{18} T^{6}_{\rm MeV}
\rho 4 \pi r^{2} dr \; {\rm erg/s}.
\end{equation} 
By setting the neutrino luminosity equal to the Compton Eddington
luminosity ($L_{\rm Edd,e^-} \approx 4 \times 10^{38}$erg s$^{-1}$; the material is still
fully ionized), we estimate the temperature at the base of the atmosphere
to be $\sim 0.4$ MeV\footnote{This critical temperature depends upon
$\alpha$ and for each atmosphere we must calculate the exact critical base
temperature.  However, the strong dependence of the neutrino luminosity on
temperature prevents significant deviation from 0.4 MeV.}.  As the
temperature drops to $\sim 0.2$ MeV, the neutrino luminosity becomes
negligible ($< 10^{-3}L_{Edd,e^-}$).  
Choosing a neutrino luminosity, we can then
iteratively derive the temperature at the surface of the neutron star
($T_{\rm NS}$) and the value of the initial accretion rate $\dot{M}_0
t_0^{\alpha}$ for a given $\alpha$.  Using this relation, we can derive
$\rho_0$, $P_0$, and $M_{\rm atm}$.  Hence, the structure of our
photon-cooling atmosphere is uniquely defined by a given $\alpha$.

Our simulations (see the appendix for a description) model both neutrino
emission and absorption as well as photon transport. We set up a range of
infall initial conditions by varying $\alpha$ and allowing the matter to
build an atmosphere around the neutron star. If the material begins to cool
significantly through neutrino emission, it compresses above a critical
density and we accrete it onto the neutron star, removing it from the
atmosphere.  The atmosphere which remains will persist until it can be
accreted via photon emission.  In this section, we will discuss the
structure of this remaining atmosphere.

Figure 2 shows the simulated fallback profiles (for $\alpha=1.5$, $\sim
30$s after the initial fallback) of density, temperature, pressure, and
entropy and the corresponding profiles of these quantities from equations
(\ref{eq:entr}, \ref{eq:pres}-\ref{eq:dens}).  The entropy gives the best
indication of the fit as it does not vary so dramatically with radius.
Figure 3 shows entropy profiles $\sim 30$s after the initial fallback for
a series of atmospheres varying $\alpha$.  Our analytic description of the
atmosphere agrees well with the simulations until the material becomes
sufficiently transparent to photons that the effects of photon diffusion
become important.

At this point in the simulations, the density and temperature structure of
the atmosphere near the base of the neutron star is well defined and we can
determine much of the internal structure of the atmosphere.  Table 1 gives
the derivations of $P_0$, $T_0$, and $\rho_0$ for a range of values for
$\alpha$ which, combined with equations (\ref{eq:pres}-\ref{eq:dens}),
defines the internal structure of the atmosphere.  We will use these
properties in \S 2.3 to discuss the the late-time evolution of the
atmospheres.  Table 1 lists both the analytic estimate of the total
enclosed mass for each atmosphere and that obtained through simulations.
For comparison with the simulations, we list simulation results when
$L_{\nu}=10^5 L_{Edd,e^-}$.  The table also includes the data for a
neutrino luminosity of $L_{Edd,e^-}$ for several values of $\alpha$.  The
accretion timescale ($t_{\rm acc}$),
\begin{equation}
t_{\rm acc}=\frac{G M_{\rm NS} M_{\rm atm}}{r_{\rm NS}
{\rm max} (L_{\nu},L_{Edd,e^-})},
\end{equation}
increases dramatically as the atmosphere cools and can be as high as $\sim
10^6$s when the neutrino luminosity equals the Compton Eddington luminosity.

Most of the material quickly cools and accretes via neutrino emission.  
Although the cooling time is proportional to the binding energy, and 
hence mass, of the atmosphere, the strong dependence of the neutrino 
cooling on the temperature, $\propto T^9$, causes the last bit of 
material to take the longest to accrete.  The temperature at the 
base of the atmosphere $T_{0} \propto M_{0}^{1/4}$ (eqs. 10,12) 
and neutrino cooling, then, is roughly proportional to $M_{0}^{9/4}$.  
The mass at which $L_\nu = L_{\rm Edd,e^-}$ defines a sharp boundary between 
rapid neutrino cooling above and constant cooling due to photons 
below this condition.  Because the neutrino cooled phase is so rapid, 
we can estimate the total accretion timescale to be equal 
to the accretion timescale at this critical mass.

\subsection{Post-Fallback Evolution}

Returning to our discussion of the neutron star atmosphere, as the
temperature at the base of the atmosphere drops below $\sim 0.4$MeV, the
neutrino cooling becomes negligible.  At this point, the cooling of the
atmosphere depends almost entirely upon the photon heat transport.  If the
dominant opacity source were electron scattering, the luminosity of the
accreting neutron star would roughly equal the usual Compton 
Eddington luminosity.  The strong dependence of the opacity on temperature
as the temperature drops below 0.2 keV creates a narrow transition region
from electron-scattering to an opacity dominated by recombined heavy
elements, especially those of the iron group (see Figure 4). 
Thus, where the temperature drops below this value, there is a 
narrow transition region within which
the Eddington limit falls dramatically.

Figure 5 shows the luminosity profile for a simulation with $\alpha=1.1$ as
a function of time.  The region whose opacity is dominated by 
iron\footnote{In the regimes that are most important for our fallback
simulations, the results do not change significantly (30 \%) if the
fallback material were instead composed entirely of oxygen.} forms an
insulating blanket around the atmosphere, limiting its cooling rate.  At
this boundary, photon pressure halts the infall and eventually drives off
the fallback material in a wind. Almost all of the photon energy generated
below this opacity boundary, at the Compton $L_{\rm Edd,e^-}$, 
goes into ejecting
mass, doing work against its gravitational binding; in other words, the
accretion flow bifurcates, the outer layers reversing direction and forming
a wind. Because the accretion rate (from larger radii) has fallen nearly to
zero by this time, this wind meets virtually no resistance and is driven
off to ``infinity''.

Figure 6 shows mass-point trajectories for a typical simulations
($\alpha=1.1$).  Much of the matter accretes onto the neutron star, while
the rest is driven off in a wind.  The atmosphere is removed over a cooling
timescale ($\lesssim 10$\,days -- see Table 1).  As the atmosphere expands
in response to the decreased pressure of its diminishing outer layers, the
temperature decreases in deeper layers. These layers then also
recombine and, in turn, are driven off.  This process continues 
until the atmosphere either accretes onto the neutron star or is 
blown off in the accretion-driven wind.

During this late phase, the observable luminosity of the accreting neutron
star is the fraction of the photon emission which escapes conversion into
kinetic energy in the ``blanket'' -- i.e. the iron-opacity Eddington limit.
Figure 7 shows the photon luminosity at infinity throughout the accretion
process; after about 500 seconds the luminosity is seen to fall by the
expected 3 to 4 orders of magnitude. Any subsequent accretion will continue
to produce a luminosity limited to $\lesssim 10^{-3}$ of the fully-ionized
Compton Eddington limit.

We have thus established that after a very brief interval, essentially none
of the original fallback is left to accrete onto the neutron star at later
times. 

\section{Fallback Accretion 10 Years After}

We have shown in \S 2 that any of the ``late-time'' fallback 
predicted by Woosley \& Weaver (1995) is removed within a 
year of the supernova explosion.  But what about 10 years later 
for the case of SN 1987A?  Present calculations of fallback 
(e.g. Woosley \& Weaver 1995) can tell us little about very low 
rates of accretion. The Compton Eddington accretion limit is only 
$10^{-8}M_\odot\,$y$^{-1}$; such a small value is far below the resolution 
limit of present-day hydrodynamic calculations (being much less than a single 
zone per year in the current fallback simulations). 
It is not at all unreasonable to expect that, 
while the accretion rate declines rapidly to such small values, 
it may remain at such values for periods of up to
many years. 

For example, years after the explosion, the decay of $^{56}$Ni in the 
supernova may decelerate some of the innermost material to the point 
where it becomes once again gravitationally bound to the neutron star.
The energy available from radioactive decay of $^{56}$Ni 
to $^{56}$Fe ($9.3\times 10^{16}$erg g$^{-1}$) corresponds to a
velocity increment of $4300$\,km\,s$^{-1}$ and this energy will both 
accelerate the outer mass {\it and} decelerate the inner region of the
supernova ejecta.  

By extrapolating the results from Woosley \& Weaver (1995), we can 
estimate the accretion rate, and hence potential energy from 
accretion released from fallback years after the supernova event.
If we take an arbitrary ejecta velocity of the inner 
zone from the simulations by Woosley \& Weaver ($\sim 10^{-3} M_\odot$) 
to be $100$km s$^{-1}$, after 1000 days its radius is $\sim 10^{15}$\,cm.  
At these times, the density of this inner material has decreased to 
$\sim 10^{-15}$g\,cm$^{-3}$.  Due to the homologous outflow which 
characterizes supernova explosions, the outer material has 
expanded at even higher velocities and the density decreases 
sharply with radius.  If this inner zone were to decelerate 
and become bound to the neutron star, it would accrete roughly 
over a free fall time, and the accretion rate ($\dot{M}=
M_{\rm inner \; zone}/t_{\rm free \; fall}$) would be $\sim 10^{-13} M_\odot 
{\rm s^{-1}}$.  The potential energy released during this time 
yields a luminosity $\sim 3\times 10^{33} {\rm erg s^{-1}}$, 
4 orders of magnitude below the detection threshold.  

However, we can only make rough estimates given the current 
resolution of the simulations and we can not preclude 
accretion rates of $\sim 10^{-9} M_\odot {\rm s^{-1}}$. 
Such small accretion rates, if all their accretion energy were 
converted to photons, would be detectable above the standard decay 
light curve of SN 1987A.  In this section, we direct our attention 
to the possibility that accretion rates as high as 
$\sim 10^{-9} M_\odot {\rm s^{-1}}$ might occur after 10 years.  

Since the composition of this late-time fallback will still be
representative of the innermost layers of the ejecta, it will still have an
opacity which is dominated by heavy elements and thus have an Eddington
limit far below the Compton value.  Just as in \S 2 where the 
iron opacity converted most of the potential energy of the 
accreting material into kinetic energy of an accretion driven 
wind, we find that even after 10 years, line-driven opacity 
will limit the accretion rate.  However, we now move into regimes 
where the SESAME opacity table is no longer accurate and must, 
in this  section, discuss the fundamentals of line-driven opacity and 
apply their consequences on the emission from this additional 
fallback.

\subsection{Line-Driven Opacity}

The increase in mean opacity above the Thomson opacity is a 
simple consequence of atomic physics.
The Thomas-Rieche-Kuhn sum rule requires that the sum over all the
oscillator strengths $f_{ij}$ of transitions arising from a given atomic
level equals the number of electrons which participate in giving rise to
the transitions from that level. Most often, this just means that the sum
of all the $f_{ij}$ over all transitions arising from a given level equals
unity.  Since the sum over all of the populations of all levels in an ion must
equal the density of the ion, the sum of all the oscillator
strength in an ion must be equal to the number of bound electrons. How,
then, does having a large number of transitions in the spectrum of an ion
increase the radiation force above that for Thomson scattering?

The answer is that the ``oscillator strength'' of a free electron is spread
out over (nearly) all energies, while the effect of atomic physics is to
concentrate the oscillator strength within a relatively narrow range of
energies, characteristically within a factor of ten of the ground-state
ionization threshold. If this range of energies coincides with that of the
radiation field (so that the Rosseland mean weights it most heavily), the
mean opacity can then be greatly increased.  Further, if the matter and
radiation temperatures are not too dissimilar, the dominant ionization
stages will always occur such that the peak of the radiation field is close to
the typical energy of the most important transitions.

Thus, as soon as the material in the atmosphere begins to recombine, it's
opacity rises sharply. This effect is strengthened at high densities by
pressure broadening.  The strong lines in the spectrum are nearly black 
at their cores, and pressure broadening allows each line to absorb 
a greater fraction of the spectral bandwidth.  At densities below 
which pressure broadening has a significant effect, the opacity can 
still be increased by the presence of a velocity gradient in the flow 
-- the spread of co-moving frame energies ``seen'' by a photon with 
fixed observer-frame energy again allows a single optically thick line 
to absorb greater fraction of the total bandwidth.

The effect of pressure broadening is included in the high-density opacities
we have employed from the SESAME code (Magee 1993). At densities below
$\sim10^{-11}$g\,cm$^{-3}$ and temperatures below $\sim 1$eV these opacities
become unreliable due to incompleteness in the atomic physics
calculations. At significantly higher densities, the effects of a velocity
gradient, which are not included in the SESAME data, have already become
important. We therefore take the larger of the results from the SESAME
tables or the ``expansion opacity'' descibed in the following sections.

In the SESAME opacities employed above, for temperatures near 0.1 MeV
the main contribution arises from bound-free transitions and lines
significantly broadened by the high density. Further out in the flow, at
lower temperatures and densities, the ions become more recombined and lines
begin to dominate the opacity. Because of the density of lines in
iron-group elements, the opacity is strongly increased by a velocity
gradient -- the larger the velocity gradient ($\propto t^{-1}$ for
homologous expansion), the greater its opacity and the lower the Eddington
limit. Even for Fe II at nebular temperatures, the Eddington limit in the
inner layers of the ejecta is much reduced below the electron-scattering
value.  Thus, even for very cool material far from the neutron star, 
matter is accelerated outward by a luminosity below observational limits.
We argue that any possible accretion rate is thus severly limited.

\subsection{Heating and Momentum Transfer}

There are two ways in which an increased opacity can affect an accretion
flow.  The first is by increasing the heating rate in the high-opacity
material. In the diffusion limit, if the flow is to sustain a given
luminosity it must achieve a sufficient temperature gradient to overcome
the effect of the opacity by heating the inner parts of the high-opacity
region.

As we have shown in \S 2, the entropy profile that develops for these
fallback atmospheres is convectively stable (see Figures 2,3).  As long as
the accretion rate falls off rapidly ($\alpha > 1$), the atmosphere will not
be convectively unstable. However, in the innermost parts of the 
high-opacity region, the increased heating rate raises the entropy and 
initiates convection.  In our simulations of the initial fallback, 
this entropy gain was limited to the inner zone of the high-opacity 
region and the resultant convective region would lie in a very 
narrow radial space.  

Even if the boundary between the high- and low-opacity regions is disrupted
by an instability, beyond the unstable region the radiation field can {\em
still} not be greater than the appropropriate Eddington limit.  The
accretion flow will be stopped by radiation pressure at the {\em largest}
radius at which the radiation force exceeds gravity. As one moves outward
in radius, the flow can be expected to become more uniform and the
radiation field more spherically symmetric, so that at such radii the usual
Eddington argument still applies. Thus, the detailed nature of the flow
near the neutron star seems not to be crucial to our argument.

\subsection{The Eddington Limit for a Line-Dominated Opacity}

An increased opacity can also increase the rate of momentum transport from
the radiation field to the gas.  The spherically-symmetric gas momentum
equation in the co-moving frame is, to $O(v/c)$,
\begin{equation}
\rho{{Dv}\over{Dt}} = 
-{{G M(r) \rho}\over{r^2} } - {{\partial p}\over{\partial r}}
+ {1\over c}\int_0^\infty F_\nu \chi_\nu d\nu
\label{eq:radmom}
\end{equation}
The last term is the radiation force per unit volume (or momentum flux per
unit length), where $\chi_\nu$ is the the extinction coefficient (units:
length$^{-1}$).

In the usual treatment of the Eddington limit, the opacity is taken to be
due only to free electrons, $\chi_{e^-} = \rho N_A f
\sigma_e/\overline{A}$, where $f$ is the mean number of free electrons per
ion, $\bar{A}$ is the mean atomic weight of the material, and $\sigma_e$
is the Compton cross section at the flux-weighted mean energy of the
radiation field. Assuming a radially-streaming radiation field with total
flux $F$, we have the radiation force
\begin{equation}
\phi_{e^-} = {{F \rho N_A f \sigma_e}\over{c\overline{A}}}
\end{equation}
or $\phi_{e^-} = 1.3\times 10^{-11} \rho F f/\overline{A}$ in cgs units in
the low-energy (Thomson) limit.

If we set $F=L/4\pi r^2$ and ignore the gas pressure gradient, equation
(\ref{eq:radmom}) gives us the usual Eddington limit
\begin{equation}
L_{\rm Edd,e^{-}} = 1.25\times 10^{38} \left({{M}\over{M_\odot}}\right)
{{\overline{A}}\over{f}}{\rm erg s}^{-1}
\end{equation}

This limit is an absolute upper bound on the photon luminosity generated by
accretion in a steady state, spherically symmetric flow. 
Above this limit the radiation force on the free electrons alone
exceeds gravity. Without neutrino cooling, accretion cannot take place at
rates much larger than this value. Even where neutrino cooling is
important, the excess photon luminosity above the Eddington limit will
drive mass outflow, and the photon luminosity at infinity will fall near or
below the limit.  For once-ionized material dominated by iron-goup elements, 
as expected for the inner layers of the ejecta, we have $L_{\rm Edd,e^-}
\sim 7\times 10^{39} (M/M_\odot)$erg s$^{-1}$, or about $10^{40}$erg s$^{-1}$.

We have already discussed the increased opacity at high densities
immediately after fallback. In the supernova ejecta at late times the effective
opacity will also be much larger at nebular temperatures
and the Eddington limit will again be correspondingly much lower.
Eastman and Pinto (1993) give an expression for the opacity of a large
number of lines (the ``expansion opacity'') in the Sobolev limit as
\begin{equation}
\chi_\nu = \nu {\beta \over r}
{{\sum_j \int_0^1 (1+Q\mu^2)\left\{ 1 - e^{-\tau_j(\mu)}\right\}d\mu}\over
{\Delta\nu}},
\end{equation}
where for an outflow the sum extends over all lines in the interval
$[\nu,\nu+\Delta\nu]$. $\mu$ is the direction cosine from the radial
direction, and $\tau_j(\mu)$ is the Sobolev optical depth of line
$j$ in the direction $\mu$,
\begin{equation}
\tau_{lu}(\mu) = {h\over{4\pi}}{{n_l B_{lu} - n_u B_{ul}}\over{|\partial \beta/
\partial l|}},
\end{equation}
or, neglecting stimulated emission,
\begin{equation}
\tau_{lu}(\mu) = {{\pi e^2}\over{m_e c}} \nu_{lu}^{-1} f_{lu} n_l
\left|{{\partial \beta}\over{\partial l}}\right|^{-1},
\end{equation}
where $f_{lu}$ is the oscillator strength of the transition and the
directional derivative is
\begin{equation}
{{\partial \beta}\over{\partial l}} = {{\partial \beta}\over{\partial r}}
\left(1+Q\mu^2\right),
\end{equation}
with
\begin{equation}
Q = {{\partial ln\beta}\over{\partial ln\,r}} - 1.
\end{equation}
$Q$ is zero for homologous flow (throughout most of the ejecta in the
supernova), and takes on a value of -3/2 for free infall.

Following the discussion in Eastman \& Pinto (1993), the integral over
angle
\begin{equation}
I_j = \int_0^1 (1+Q\mu^2)\left\{ 1 - e^{-\tau_j(\mu)}\right\}d\mu
\end{equation}
has the limits $(1 + Q/3)$ for $\tau_j(\mu=0)>>1$ and $\tau(1+\tau Q/3)$ for
$\tau_j(\mu=0)<<1$. We will approximate the integral as
\begin{equation}
I_j = \tau^*\left( 1 + {{\tau^*_j Q}\over 3}\right)
\end{equation}
with $\tau^*_j = min(\tau_j(\mu=0),1)$. This gives us the opacity
\begin{equation}
\chi_\nu = \nu {\beta \over r}
{{\sum_j \tau^*\left( 1 + \tau^*_j Q/3\right)}\over{\Delta\nu}}.
\end{equation}

The value of this opacity clearly depends upon knowing the spectral density
of line transitions: the number of lines per unit frequency.  While line
lists are available from which we can determine the spectral density, most
such lists are seriously incomplete, especially in the context of
heavy-element-rich supernova ejecta.  As a purely illustrative example, we
have employed the line list of Kurucz (1991) which lists roughly 171,000
lines of Fe and Co with lower-level energies of $10^5$cm$^{-1}$ or less. The
true value of the opacity will be rather larger due to the incompleteness
of the list.

We have assumed an LTE equation of state at a temperature of 5000K.  Such a
temperature is typical of models for supernova ejecta at times later than a
few hundred days. The matter is assumed to be expanding homologously
($\partial v/\partial r = v/r = 1/t$) and has a density scaled from the
inner zones of hydrodynamic simulations. We have chosen a density range
with $\rho = 10^{-8} - 10^{-6}$g\,cm$^{-3}$ 
as being typical.  Figure 8 shows the
computed Eddington limits from this model. At present (10 years past
explosion), the Eddington limit is near if not below the observational
limit. Calculations, such as those of Woosley \& Weaver (1995), show a
density near $5\times 10^{-7}$g\,cm$^{-3}$ 
(scaled to one day) in their innermost
zones, but the density profile from these simulations falls off very
strongly with decreasing radius and the mass resolution of the
innermost zone only $\sim 10^{-3}M_\odot$ which makes it difficult 
to determine the density profile. Unfortunately, without far more 
detailed simulations of the dynamics of the innermost portion of the 
ejecta, a firm prediction of whether the ejecta Eddington limit is
below the observed luminosity is impossible.

If the neutron star is given a significant impulse (``kick'') during the
explosion, it will find itself moving with the flow at the same velocity.
Thus, the material surrounding it will have zero velocity in its own frame.
Because the velocity gradient in homologous expansion is isotropic, the
flow will still be isotropically outward from the neutron star, and the
arguments of this section will remain valid.

\section{The remnant of SN 1987a}
We have shown that virtually all of the late-time fallback of material onto
a nascent neutron star which occurs in most supernova simulations (Woosley
\& Weaver 1995) accretes rapidly onto the neutron star.  However a small
remnant of this material ($\sim 10^{-11}-10^{-9}M_\odot$) remains on the
neutron star and cools due to photon emission.  Due to the high opacity of
recombined iron, the atmosphere emits at a rate much below Compton
Eddington ($\sim 10^{-3}L_{\rm Edd,e^-}$), with most of the photon energy
driving a wind from the neutron star.  The accretion times of the
atmospheres are $\lesssim 0.1$\,yr.  Any further fallback will be limited to
the iron-opacity Eddington accretion rate.  In the case of 1987a, this
means that, although we would not expect to see any luminosity from the
fallback.  

If no mechanism drove further fallback onto the neutron star 
(e.g. the decay of $^{56}$Ni), then the neutron star remnant of SN
1987A should now have no obscuring atmosphere.  If this did indeed 
occur, and if the neutron star is rotating rapidly and has a 
strong magnetic field (such as the Crab Pulsar), then we should 
observe it as a pulsar.  Only when all of these ``ifs'' are satisfied 
must we, as Bethe \& Brown (1995) have suggested, conclude that 
the neutron star has indeed collapsed into a black hole.

\acknowledgements We acknowledge the early work of Kaiyou Chen, whose 
unpublished research first recognized the importance of the opacity 
of recombined iron.  We are grateful to Stan Woosley for access to fallback
models and for many discussions related to fallback and fallback rates.  We
would also like to thank Roger Chevalier, Adam Burrows and Willy Benz for
many useful discussions on this topic.  The work of C. Fryer was supported
by the NSF (AST 94-17161) and by NASA (NAG5 2843 and MIT SC A 292701) and
of P. Pinto by the NSF (CAREER grant AST9501634) and by NASA (NAG\,5-2798).
P. Pinto gratefully acknowledges support from the Research Corporation
though a Cottrell Scholarship.  S. Colgate acknowledges the support 
of LDRD funds through DOE and UC.  

\appendix
\section{Code Description} For these simulations, we have used
the one-dimensional lagrangian code tested in several previous papers
(Herant et al. 1994, Fryer et al. 1996).  The equation of state and the
neutrino transport and emission/absorption processes are described in
detail in Herant et al. (1994). In this appendix, we describe the physics
added to the code to simulate late-time fallback onto neutron stars.  We
begin by describing our initial conditions and how we ``accreted'' material
onto the neutron star.  We end with a discussion of our photon transport
and opacity calculations.

We begin with free-fall initial conditions where the accretion rate onto
the neutron star is $10^5M_\odot\,$y$^{-1}$.  
This rate declines proportionally to
$t^{-\alpha}$ where $\alpha$ is a free parameter.  We use 500 zones to
model the entire fallback material from 10\,km - $6 \times 10^5$\,km.  These
zones vary in size from $\sim 1$\,km near the surface of the neutron star to
$\sim 10^4$\,km near the edges of our simulation.  In our lagrangian code,
these sizes vary during the course of the simulation, but not noticeably.
As the cell next to the surface of the neutron star cools and its density
increases beyond a critical density ($\rho_{\rm crit}$), that cell is
accreted.  The value for $\rho_{\rm crit}$ is gradually decreased 
($10^{11}-10^{5}$\,g\,cm$^{-3}$) during
the simulation after the initial structure of the atmosphere is defined,
allowing us to follow the evolution of the atmosphere to later times
($10^3-10^4$\,s).

For photon opacity, we use the SESAME opacity data (Magee 1993) for the
high density/high temperature regimes.  Below a density of
$10^{-11}$g\,cm$^{-3}$ or a temperature below 1eV, the SESAME data is
incomplete.  In this regime, we use the approximation for the expansion
opacity at the Sobolev limit by Pinto (1997):
\begin{equation}
\chi = \frac{D \beta}{r}
\tau^{*} \left( 1 + \frac{\tau^{*} Q}{3} \right),
\end{equation}
$D$ is the number of lines per frequency bin, $\beta =$max(material
velocity,sound speed)/(speed of light), $\tau^{*}=$min(1,$10^{13} \rho$
cm$^3$/g), and $Q=\frac{\partial ln \beta}{\partial ln r} -1$.  The
boundary between these two regimes does not reflect a smooth transition;
the low-density expansion opacity approximation is lower than the
predicted opacity from the SESAME data.  Our low-density opacity is likeley
to be an underestimate, implying that our calculated photon luminosities
are higher than the actual values.  We have also run the same simulation
assuming that the opacity remains closer to the SESAME data in the
low-density regime and the photon luminosity did not decrease significantly
as most of the flux is coming from the high-density regimes where we use
the SESAME data.

The photon transport is modeled using the Levermore-Pomraning
flux limiter (Levermore \& Pomraning 1981).   Our assumption that
the photons are emitted  in local thermodynamic equilibrium from
each cell holds for  most of atmosphere where the cell sizes are
much longer than  a few mean-free paths.  In fact, if anything,
the mean photon  energy would be greater in the outer part of the
atmosphere  where the atmosphere becomes optically thin, leading
to an  opacity which is higher than predicted by our simulations,
implying once again, that these accreting neutron stars  are less
luminous than we have predicted.

\begin{deluxetable}{lccccccc}
\tablewidth{42pc}
\tablecaption{Atmosphere Structure}
\tablehead{ \colhead{$\alpha$} & \colhead{$S_0$}  &
\colhead{$P_0$} & \colhead{$T_0$}  & \colhead{$\rho_0$} &
\colhead{$M_{\rm atm}$}  & \colhead{$M_{\rm atm}^{\rm sim}$} &
\colhead{$t_{\rm acc}$}\\ & \colhead{${\rm (k_B/nucleon)}$} &
\colhead{${\rm (g/cm/s^2)}$}  & \colhead{$(\rm K)$} &
\colhead{${\rm (g/cm^3)}$} &
\colhead{$(M_{\odot})$} & \colhead{$(M_{\odot})$} & ($10^3$ s)}

\startdata
\multicolumn{7}{c}{$L_{\nu}=10^5 L_{\rm Edd}$} \nl & & & & & & \nl
$1.05$ & $115$ & $1.01 \times 10^{26}$ & $1.41 \times 10^{10}$ &
$3.02 \times 10^{6}$ & $2.0 \times 10^{-7}$ & $2.1 \times
10^{-7}$ & 2\nl
$1.1$ & $115$ & $1.02 \times 10^{26}$ & $1.42 \times 10^{10}$ &
$3.05 \times 10^{6}$ & $1.2 \times 10^{-7}$ & $1.8 \times
10^{-7}$ & 1 \nl
$1.2$ & $115$ & $1.04 \times 10^{26}$ & $1.43 \times 10^{10}$ &
$3.12 \times 10^{6}$ & $6.5 \times 10^{-8}$ & $8.4 \times
10^{-8}$ & 0.6 \nl
$1.5$ & $113$ & $1.11 \times 10^{26}$ & $1.45 \times 10^{10}$ &
$3.32 \times 10^{6}$ & $2.8 \times 10^{-8}$ & $4.4 \times
10^{-8}$ & 0.3\nl
$2.0$ & $111$ & $1.21 \times 10^{26}$ & $1.48 \times 10^{10}$ &
$3.62 \times 10^{6}$ & $1.5 \times 10^{-8}$ & - & 0.1 \nl
$3.0$ & $107$ & $1.37 \times 10^{26}$ & $1.53 \times 10^{10}$ &
$4.12 \times 10^{6}$ & $8.6 \times 10^{-9}$ & - & 0.08 \nl
$5.0$ & $103$ & $1.64 \times 10^{26}$ & $1.60 \times 10^{10}$ &
$4.92 \times 10^{6}$ & $5.1 \times 10^{-9}$ & - & 0.05 \nl & & &
& & & \nl
\multicolumn{8}{c}{$L_{\nu}=L_{\rm Edd}$}  \nl & & & & & & \nl
$1.05$ & $365$ & $1.01 \times 10^{24}$ & $4.47 \times 10^{9}$ &
$3.02 \times 10^{4}$ & $2.0 \times 10^{-9}$ & - & 2000 \nl
$1.1$ & $365$ & $1.02 \times 10^{24}$ & $4.48 \times 10^{9}$ &
$3.05 \times 10^{4}$ & $1.2 \times 10^{-9}$ & - & 1000 \nl
$1.2$ & $363$ & $1.04 \times 10^{24}$ & $4.51 \times 10^{9}$ &
$3.12 \times 10^{4}$ & $6.5 \times 10^{-10}$ & - & 600 \nl
$1.5$ & $358$ & $1.11 \times 10^{24}$ & $4.58 \times 10^{9}$ &
$3.32 \times 10^{4}$ & $2.8 \times 10^{-10}$ & - & 300 \nl
$2.0$ & $350$ & $1.21 \times 10^{24}$ & $4.68 \times 10^{9}$ &
$3.62 \times 10^{4}$ & $1.5 \times 10^{-10}$ & - & 100 \nl
$3.0$ & $339$ & $1.37 \times 10^{24}$ & $4.83 \times 10^{9}$ &
$4.12 \times 10^{4}$ & $8.6 \times 10^{-11}$ & - & 80 \nl
$5.0$ & $324$ & $1.64 \times 10^{24}$ & $5.05 \times 10^{9}$ &
$4.92 \times 10^{4}$ & $5.2 \times 10^{-11}$ & - & 50 \nl

\enddata
\end{deluxetable}


\clearpage

\begin{figure}
\plotfiddle{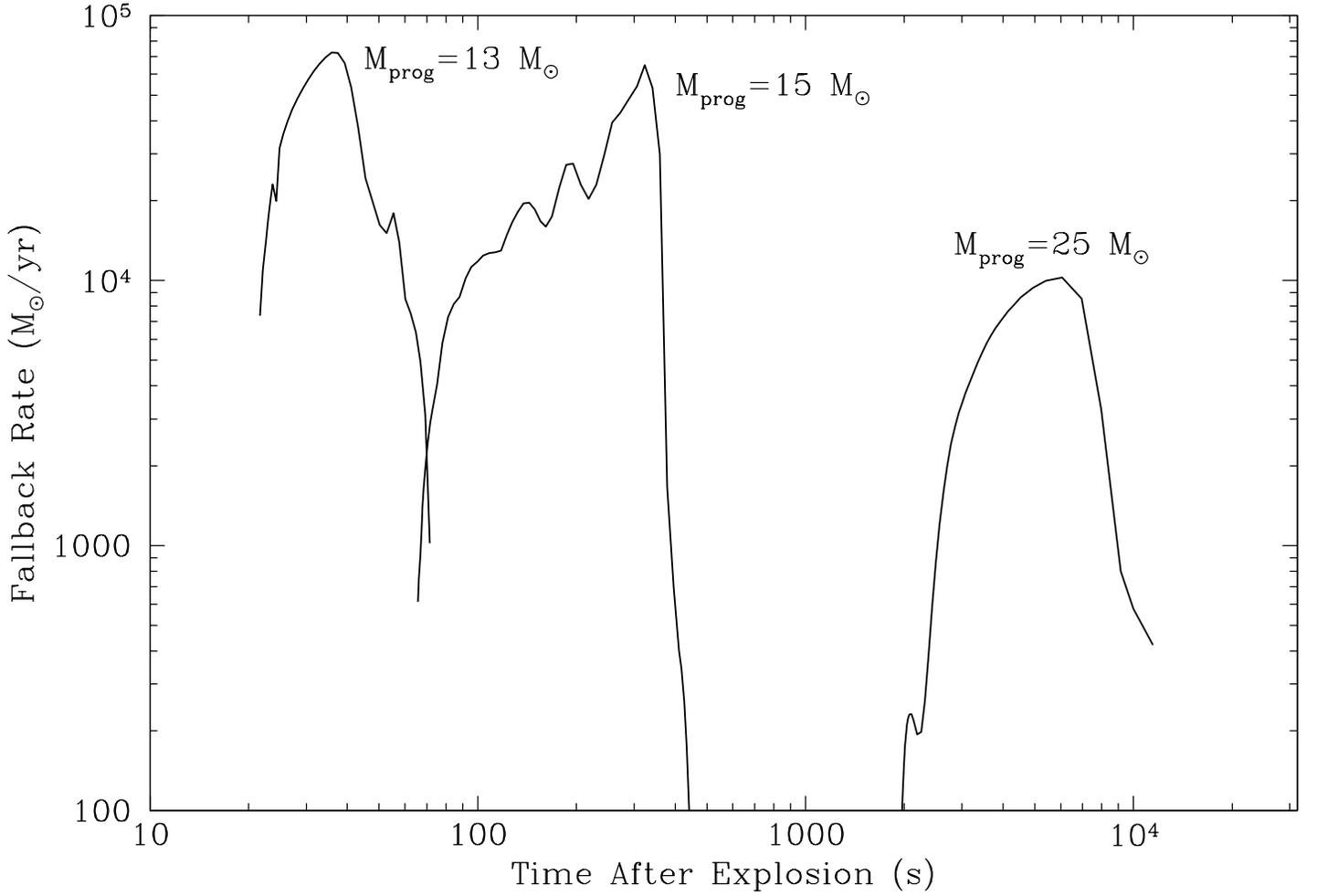}{7in}{-90}{70}{70}{-280}{520}
\caption{Estimated accretion rate vs. time using  the models of
Woosley \& Weaver (1995).  For all  models, the accretion rate
exceeds $\sim 10^4  M_{\odot}{\rm yr}^{-1}$ before dropping
dramatically.   Most of this material will accrete via neutrino
cooling.  However, the last fraction of it  ($\lesssim 10^{-9}
M_{\odot}$) will cool due  to photon emission.}
\end{figure}

\begin{figure}
\plotfiddle{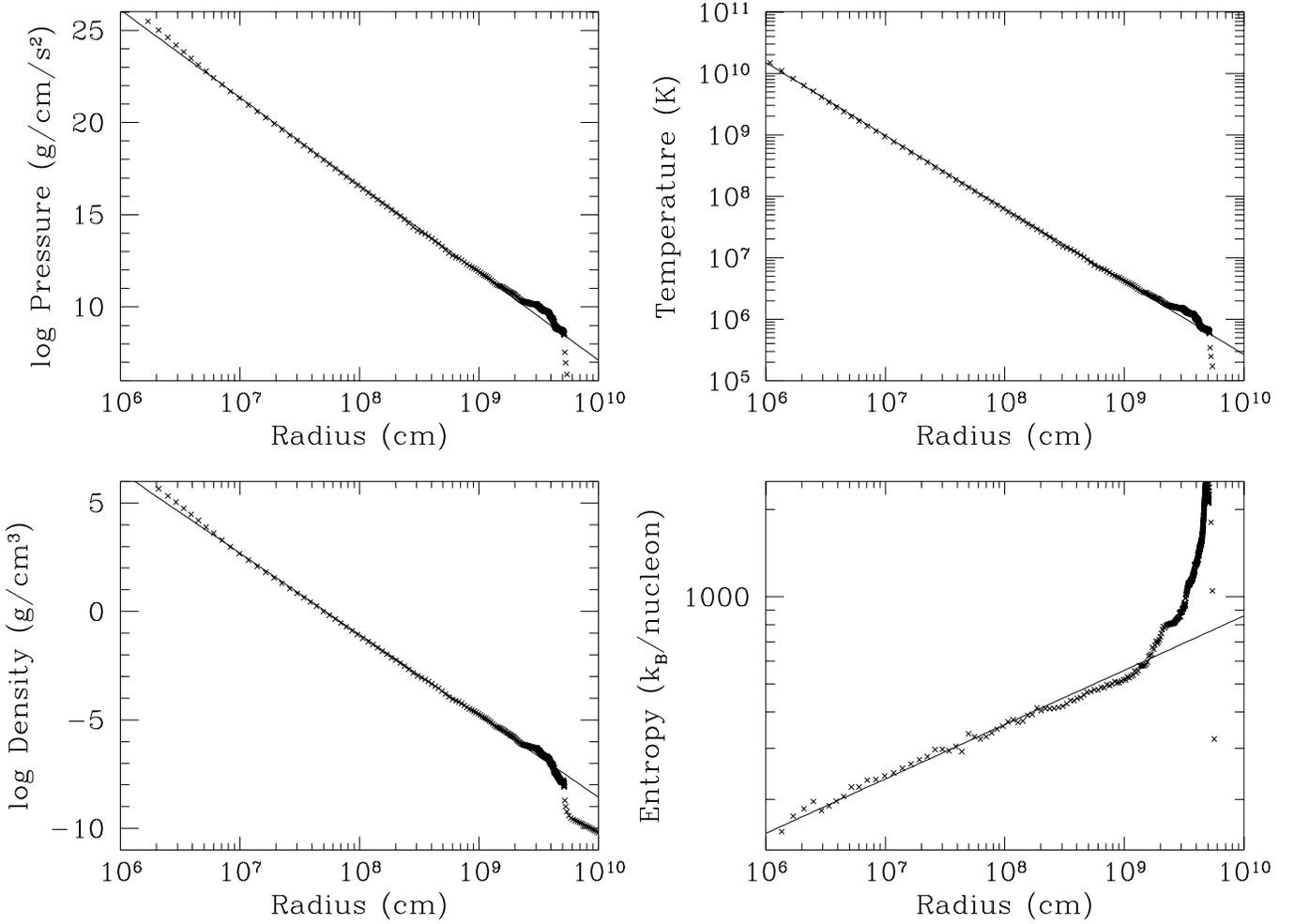}{7in}{-90}{70}{70}{-280}{520}
\caption{Comparison of simulations with the analytic  derivation
(Pressure, Density, Temperature, and  Entropy) for an
$\alpha=1.5$ model.  The analytic  model matches the simulation
well until photons are  no longer ``trapped'' in the flow.  This
occurs at  high enough radii that it does not effect the mass
estimate for the atmosphere.  Note that the Entropy  is the most
sensitive variable for comparison.}
\end{figure}

\begin{figure}
\plotfiddle{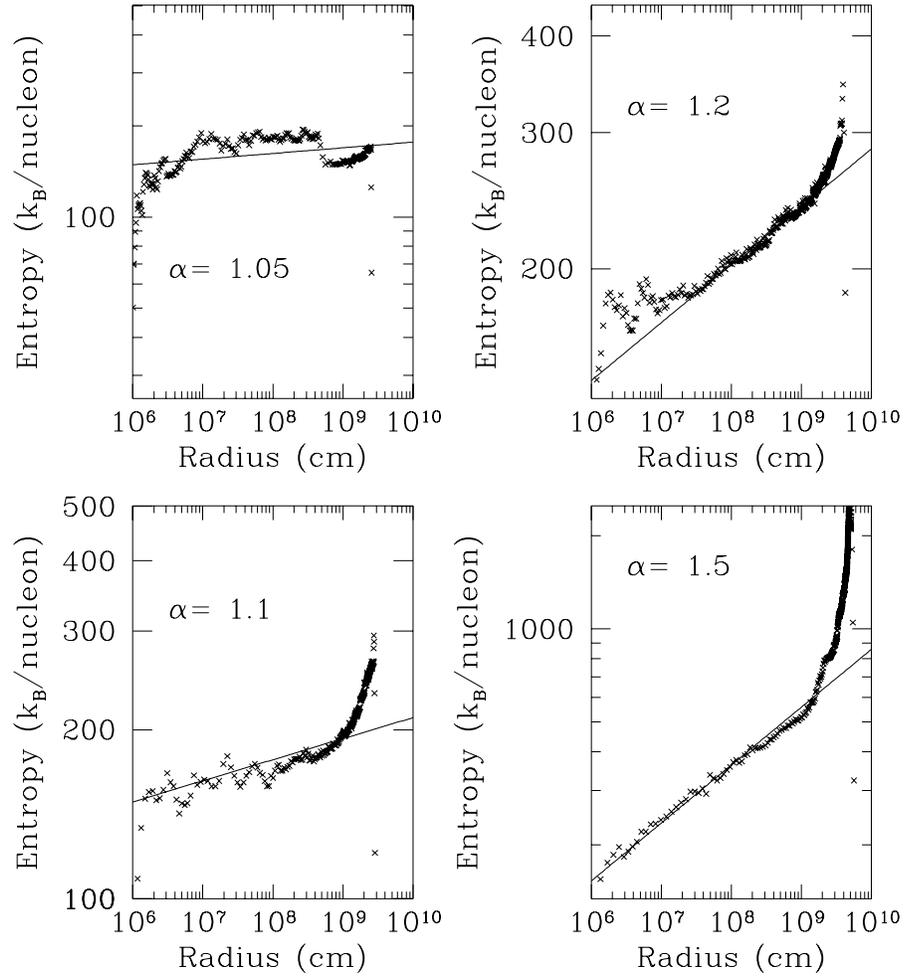}{7in}{-90}{70}{70}{-180}{520}
\caption{Entropy comparisons of simulations with analytic
derviations for a range of $\alpha$'s.  Except at  large radii
where photons are no longer trapped,  the analytic model matches
the simulations for all  values of $\alpha>1$.}
\end{figure}

\begin{figure}
\plotfiddle{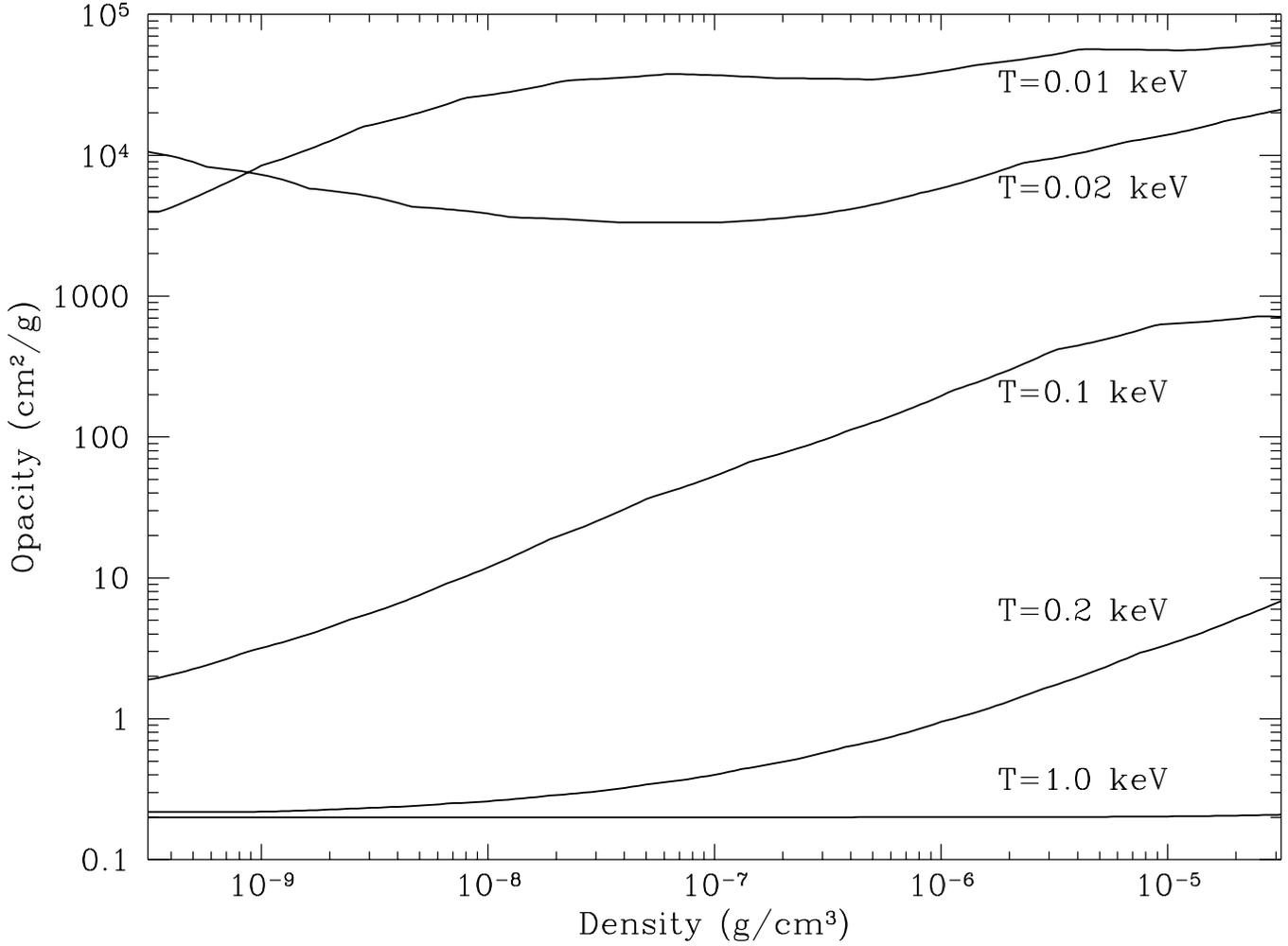}{7in}{-90}{70}{70}{-280}{520}
\caption{The iron opacity for a range of densities.   The
different lines are for different temperatures.   Note that as
the temperature drops from 0.2\,keV to  0.1\,keV, the opacity
increases by $\sim 2.5$  magnitudes.  This jump creates the sharp
transition from the roughly electron-scattering  opacity regime
to the regime dominated by recombined  iron.}
\end{figure}

\begin{figure}
\plotfiddle{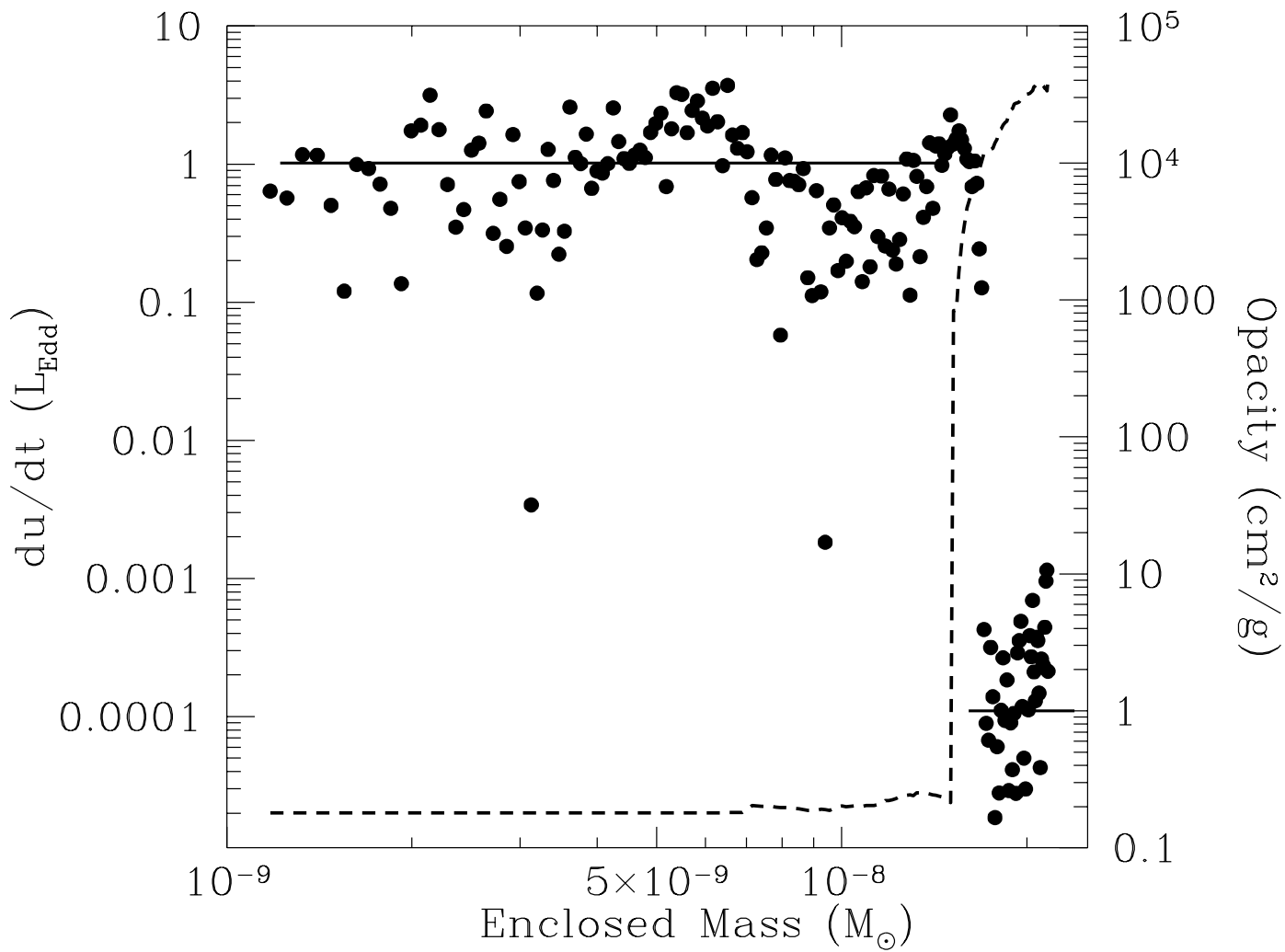}{7in}{-90}{70}{70}{-280}{520}
\caption{Luminosity (dots and solid line)  and opacity (dotted
line) vs. radius in  the neutron star atmosphere.  The dots mark
each cell in the simulation.  (The scatter exists  because the
diffusion is not in strict equilibrium  and the diffusion varies 
slightly each time step).   The net effect is that the inner
atmosphere, whose  opacity source is electron scattering has a
luminosity  roughly equal to the Eddington Luminosity.  When  the
opacity increases dramatically due to recombined  iron, the
luminosity drops to roughly $10^{-4} L_{\rm Edd}$.  It is this
luminosity that we observe.  This very large difference in
luminosity is the  energy supplied to the binding energy of the
wind.  The sharp transition in opacity is due to the sensitivity
of  the opacity on the temperature (see Figure 4).}
\end{figure}

\begin{figure}
\plotfiddle{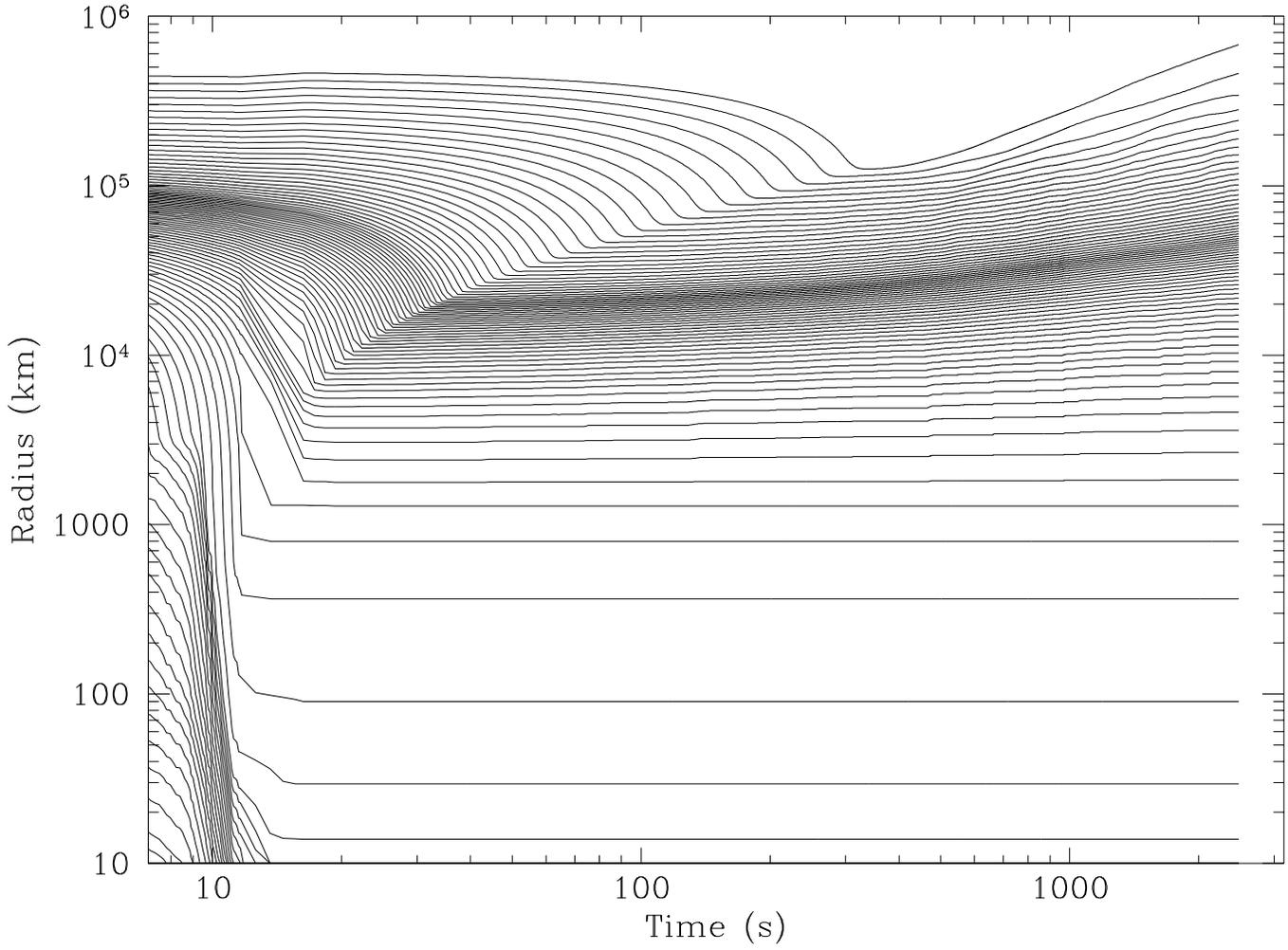}{7in}{-90}{70}{70}{-280}{520}
\caption{Mass-trajectory plot.  Note that the initial 
material is accreted rapidly as it compresses due to 
neutrino cooling and exceeds the critical density.  At 
late times, photon cooling drives a wind and 
the outer atmosphere expands.}
\end{figure}

\begin{figure}
\plotfiddle{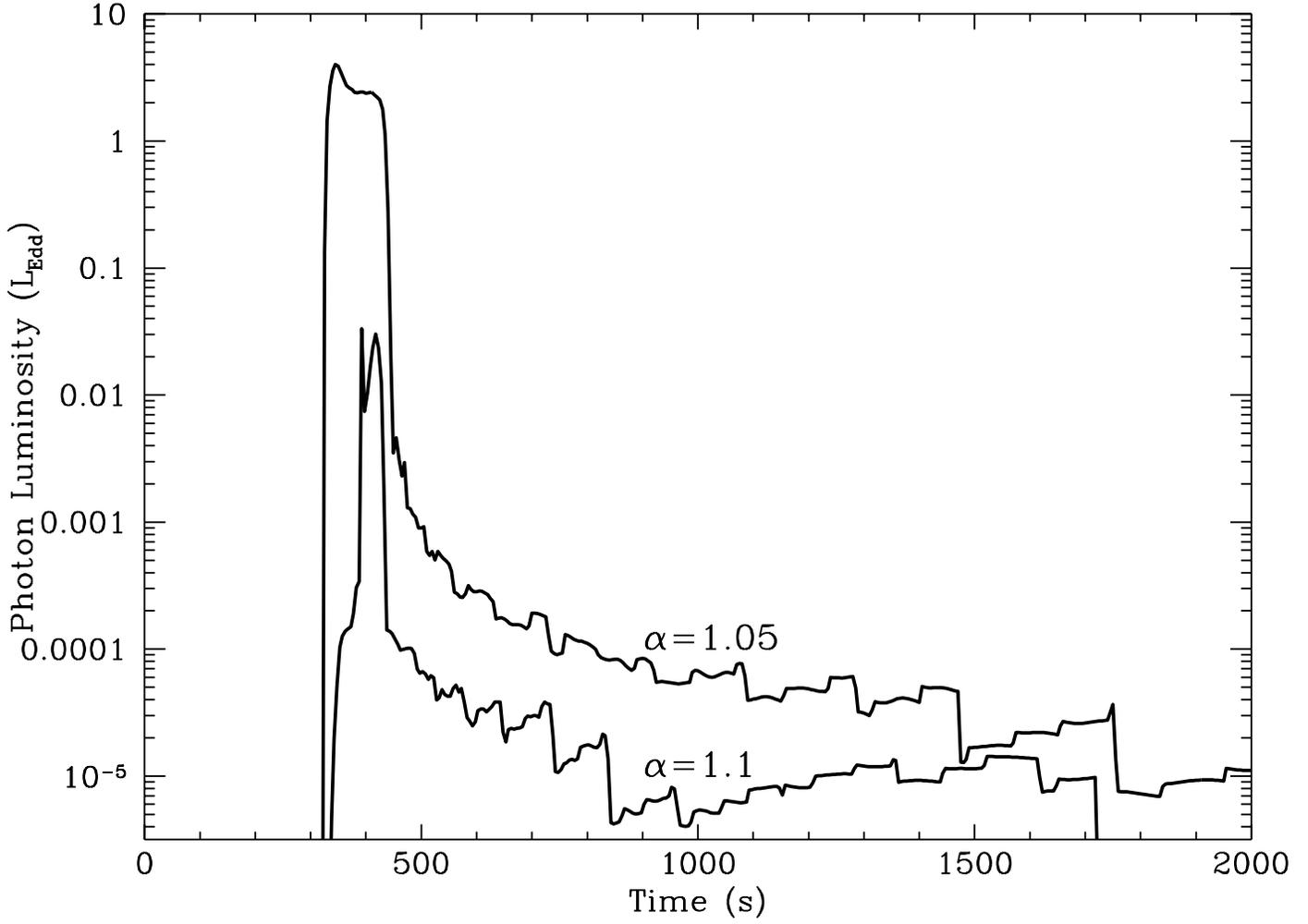}{7in}{-90}{70}{70}{-280}{520}
\caption{Luminosity of the accreting neutron  star system vs.
time.  Note that the luminosity  quickly drops to roughly
$10^{-4} L_{\rm Edd}$  and remains low for the duration of the
simulation.   During this time, the outer atmosphere blows off in
a photon-driven wind.}
\end{figure}

\begin{figure}
\plotfiddle{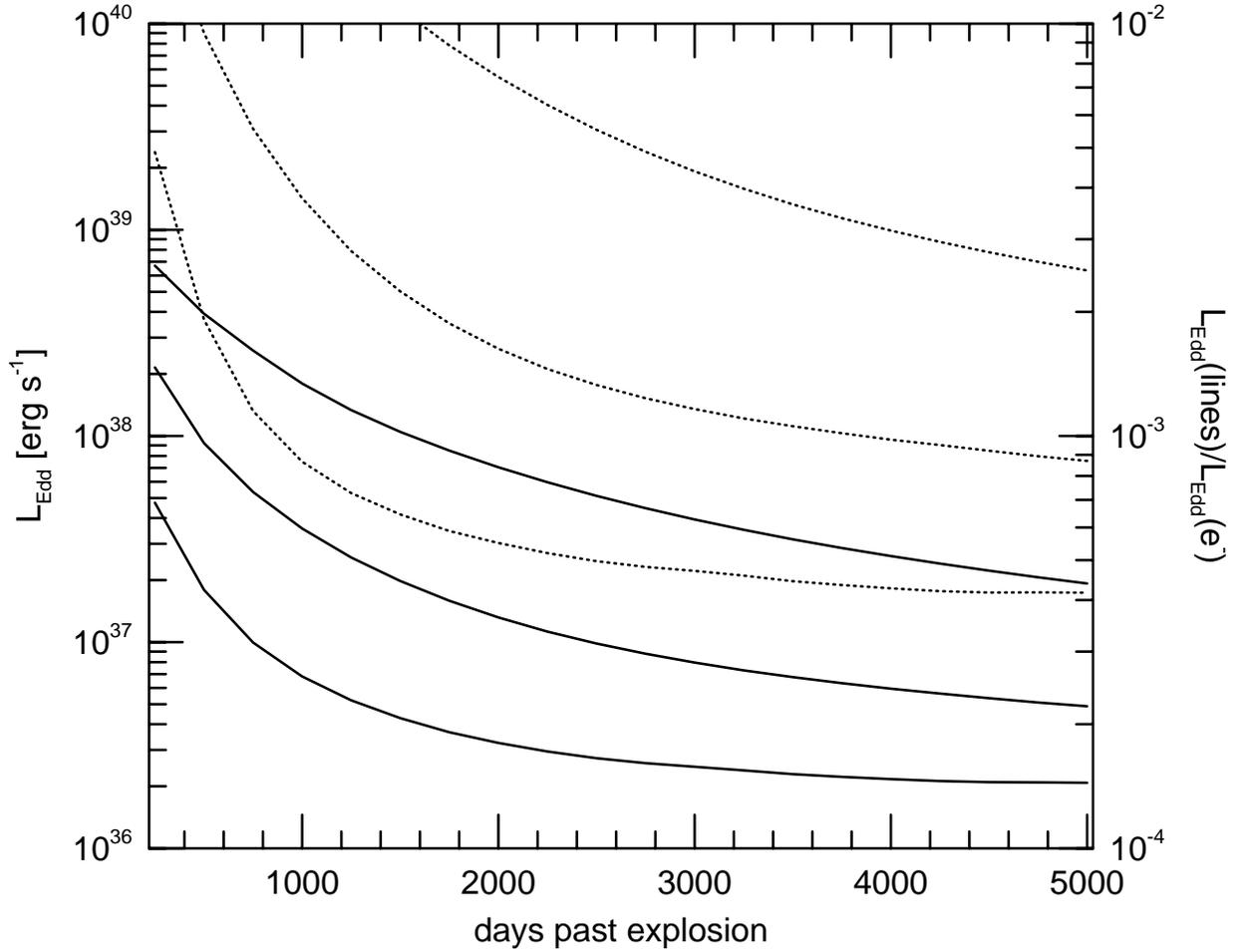}{7in}{0}{65}{65}{-250}{50}
\caption{Eddington limit including the effect of lines in a plasma composed
of Fe, Co, and Ni in solar ratios and with a density of $10^{-8}$ (lowest
curve), $10^{-7}$, and $10^{-6}$\,g\,cm$^{-3}$ 
in homologous expansion. The solid
lines show the Eddington luminosity. The dotted lines show the factor by
which the electron-scattering Eddington limit is multiplied by the addition
of line opacity.}
\end{figure}

\end{document}